\newcommand\APJ[3]{~Astrophys. J.{\bf ~#2}, #3~(#1)}
\newcommand\npb[3]{~Nucl. Phys. B{\bf ~#2}, #3~(#1)}
\newcommand\plb[3]{~Phys. Lett. B{\bf ~#2}, #3~(#1)}
\newcommand\PRL[3]{~Phys. Rev. Lett.{\bf ~#2}, #3~(#1)}
\newcommand\PRD[3]{~Phys. Rev. D{\bf ~#2}, #3~(#1)}

\documentclass[twocolumn,    
               showpacs,    
               preprintnumbers,
               aps,       
               prd,       
               a4paper,      
               superscriptaddress,
               nofootinbib,    
               tightenlines,    
               floats           
               ]{revtex4}

\usepackage{graphicx}
\usepackage{amssymb,latexsym}

\begin{document}

\title{WMAP constraints on scalar-tensor cosmology and the variation of
               the gravitational constant}

\author{Ryo Nagata}
\email{nagata@th.nao.ac.jp}
\affiliation{%
Division of Theoretical Astrophysics, National Astronomical Observatory,
2-21-1, Osawa, \\
Mitaka, Tokyo 181-8588, Japan }
\affiliation{
Department of Physics, Kyoto University, Kyoto 606-8502, Japan}
\author{Takeshi Chiba}%
\affiliation{
Department of Physics, Kyoto University, Kyoto 606-8502, Japan}
\author{Naoshi Sugiyama}
\affiliation{%
Division of Theoretical Astrophysics, National Astronomical Observatory,
2-21-1, Osawa, \\
Mitaka, Tokyo 181-8588, Japan }

\date{\today} 


\preprint{astro-ph/0311274; NAOJ-Th-Ap 2003,No.38; KUNS-1879}


\begin{abstract}

We present observational constraints on a scalar-tensor gravity theory 
by $\chi^2$ test for CMB anisotropy spectrum. 
We compare the WMAP temperature power spectrum with the harmonic 
attractor model, in which the scalar field has its harmonic 
effective potential with curvature $\beta$ in the Einstein conformal
 frame and the theory relaxes toward Einstein gravity with time. 
We found that the present value of the scalar coupling, i.e. 
the present level of deviation from Einstein gravity $(\alpha_0^2)$, 
is bounded to be smaller than $5\times 10^{-4-7\beta}$ 
($2\sigma$), and $10^{-2-7\beta}$ ($4\sigma$) for 
$0< \beta<0.45$.  This constraint is much stronger 
than the bound from the solar system experiments for 
large $\beta$ models, i.e., $\beta> 0.2$ and $0.3$ in 
$2\sigma$ and $4\sigma$ limits, respectively. 
Furthermore, within the framework of this model, the 
variation of the gravitational constant at the recombination epoch
is constrained as $|G(z=z_{rec})-G_0|/G_0 < 0.05(2\sigma)$, 
and $0.23(4\sigma)$. 
\end{abstract}

\maketitle

\section{Introduction}

The discovery of acoustic peaks on the anisotropy spectrum of
the cosmic microwave background (CMB) opened the way to measure the early 
universe. The high quality data recently provided by the Wilkinson Microwave
Anisotropy Probe (WMAP) mission \cite{benn} makes such a challenge feasible. 
The measurement of the characteristic length
of various processes at the recombination epoch 
reveals the physical process in the early universe \cite{hss}.  
Length scales, which exhibit in the angular spectrum of the CMB as a 
characteristic feature, depend on the horizon length which is controlled 
by gravity. Usually, the horizon 
length at the recombination epoch is a function of the amount of 
non-relativistic matter. The strength of gravity caused by matter 
and radiation determines how much time consumed to make the universe cool
enough to recombine hydrogens. However, the strength of gravity 
depends not only on the amount of matter but also on the coupling strength of
gravitational interaction. Thus the CMB anisotropy spectrum contains 
the information about the magnitude of the gravitational constant at the
recombination epoch. 

Constancy of physical ``constants'' is the fundamental 
issue which has long history and has attracted much interest \cite{CU}. 
The recent attempts toward unifying all elementary forces \cite{GSW} 
predict the existence of scalar fields 
whose vacuum expectation values determine the physical ``constants''. 
In such context, scalar-tensor gravity theories, whose original version was 
proposed by Jordan \cite{J} and Brans and Dicke \cite{BD}
and were extended in a more general framework later \cite{BW}, 
are the most promising candidates among the alternatives of Einstein gravity. 
In scalar-tensor theories, the coupling of a massless scalar field 
to the Ricci scalar provides a natural framework of 
realizing the time-variation of the gravitational constant
 via the dynamics of the scalar field. 

In the Jordan-Brans-Dicke theory \cite{J,BD} (hereafter, we refer
to it as the Brans-Dicke theory for simplicity) which is the simplest 
example of scalar-tensor theories, a constant coupling parameter $\omega$ is 
introduced. In the limit $\omega \rightarrow \infty$, the gravitational 
constant can not change and Einstein gravity is recovered. 
Although scalar-tensor theories including the Brans-Dicke theory are 
compatible with the Einstein gravity in several aspects, 
they have many deviations from it. Weak-field experimental 
tests in solar-system have constrained the post-Newtonian deviation 
from the Einstein gravity, $\omega > 500$
\cite{REA,CMW,robert,lebach}. 
According to some recent reports, this bound would be updated 
to several thousands \cite{eubanks,CMWGQ,ALG}. 
Such a small deviation implies that the variability of 
the background scalar field is also small. 

In more general scalar-tensor theories with non-trivial 
coupling functions $\omega(\phi)$, the small deviation is not 
the outcome of fine-tuning because the cosmological evolution drives 
$\omega$ toward infinity in the late cosmological epochs and naturally 
reduces the present observable effects of $\phi$ field \cite{DN}. 
If such attractor mechanism takes
place, the nature of gravity (the variability of the background scalar field, 
weak-field deviations, etc.) can be significantly different in 
the early universe. Hence, information on the different cosmological 
epochs may constrain such theories. A simple and natural extension 
of the Brans-Dicke theory to the attractor model is the harmonic attractor 
model in which the scalar field has a quadratic effective potential of 
a positive curvature in the Einstein conformal frame. 
The analysis of big-bang nucleosynthesis (BBN) 
in the harmonic attractor model \cite{DP} 
restricts two parameters characterizing the potential (its curvature 
$\beta$ and today's gradient $\alpha_0$). 
It is concluded that the BBN limit on the possible deviation from 
Einstein gravity ($2\omega_0+3={\alpha_0}^{-2}$) is much stronger 
than the present observational limits in large $\beta(>0.3)$ models. 
We can extract the information about the early universe 
also by the analysis of structure formation \cite{LMB}. 
The advantage of the use of the CMB fluctuations is that the physics of CMB 
is well understood and that we now have very accurate observational data. 
The trace of primordial fluctuation can be seen clearly in the CMB 
anisotropy spectrum where the information on the early universe 
up to the last scattering time is projected 
on the acoustic peaks \cite{CK,PBM,NCS}. 

We compare the CMB temperature anisotropy spectrum measured by WMAP 
with that in the scalar-tensor cosmological model mentioned above. 
The formulations of background and perturbation equations and 
the detailed explanation of physical processes can be found in our 
previous paper \cite{NCS}. 
The purpose of this work is to constrain the deviation of the 
scalar-tensor gravity from the Einstein gravity and 
to find, in the framework of 
this model, how large the variation of the gravitational constant is allowed. 

This paper is organized as follows: 
In Section II, the scalar-tensor cosmological model 
and its observational consequences are explained. 
In Section III, we describe the overview of our analysis for constraining
 the scalar-tensor coupling parameters. 
In Section IV, the models are compared with the WMAP 
data and the result of $\chi^2$ test is presented. 
Finally, some conclusions are in Section V.

\section{model and its predictions}

The action describing a general massless scalar-tensor theory is 
\begin{eqnarray}
S = \frac{1}{16\pi G_0} \int_{}^{} d^4x \hspace{0.2em}
\sqrt[]{-g} \biggl[\hspace{0.1em} \phi R - \frac{\omega(\phi)}{\phi}
(\nabla \phi)^2 \hspace{0.1em} \biggr] \nonumber \hspace{3.0em} \\
+ S_m [\hspace{0.1em} \psi,g_{\mu \nu} \hspace{0.1em}], \label{eq1}
\end{eqnarray}
where $G_0$ is the Newtonian gravitational constant measured today, 
and $\omega(\phi)$ is the dimensionless coupling parameter of the
nonminimal coupling which is a function of $\phi$.  
The last term in Eq.(\ref{eq1}) denotes the action of matter which
is a functional of the matter variable $\psi$  and the metric
$g_{\mu\nu}$. The deviation from the 
Einstein gravity depends on the asymptotic value of $\phi$ field at 
spatial infinity. According to the
cosmological attractor scenario \cite{DN}, the dynamics of the
cosmological background $\phi$ field is analogous to that of a particle 
damping its motion toward the minimum of an external potential in the 
Einstein conformal frame. 
As the generic feature of a potential near a minimum is parabolic, 
we shall study the case where the potential is quadratic. 
This setup corresponds to $\omega(\phi)$ of the following form,
\begin{eqnarray}
2\omega(\phi)+3 &=& \bigl\{ \hspace{0.1em} {\alpha_0}^2- 
\beta \ln(\phi/\phi_{0})
\hspace{0.1em}  \bigr\}^{-1}, \label{eq2}
\end{eqnarray}
where $\phi_0$, $\alpha_0$ and $\beta$ are the present value of 
background $\phi$ field, today's potential gradient and curvature, 
respectively. We consider non-negative value of $\beta$ since curvature 
near a minimum is positive. If $\beta = 0$, this model is reduced to 
the Brans-Dicke theory. Moreover, 
the model with $\alpha_0 \rightarrow 0$ and $\beta=0$ is the Einstein gravity. 
In the first post-Newtonian 
approximation, deviations from general relativity are proportional to the 
well-known Eddington (PPN) parameters as,
\begin{eqnarray}
\gamma_{\rm Edd} - 1 &=& -2 \, {\alpha_0}^2 / 
(1+{\alpha_0}^2) \, , \label{eq3} \\
\beta_{\rm Edd} - 1 &=& \frac{1}{2} \, \beta \, 
{\alpha_0}^2 / (1+{\alpha_0}^2)^2 \, . \label{eq4}
\end{eqnarray}
We see explicitly from Eqs.~(\ref{eq3}) and (\ref{eq4}) that post-Newtonian 
deviations from general relativity tend to zero with $\alpha_0$ at least as 
fast as $\alpha_0^2$. This holds true for weak-field deviations of 
arbitrary post-Newtonian order \cite{DEF}. One of the most stringent empirical
limits for PPN parameters is 
\begin{eqnarray}
-1.7 \times 10^{-3} < 4 \hspace{0.2em}  \beta_{\rm Edd} - 
\gamma_{\rm Edd} - 3 < 1.5 \times
   10^{-4} \hspace{0.4em} (1\sigma) \hspace{0.1em}, \label{eq5}
\end{eqnarray}
which is obtained by the Lunar Laser Ranging experiment \cite{WND}. 
In the framework of the present model, this translates into
\begin{eqnarray}
{\alpha_0}^2 < 1.5 \times 10^{-4} / (\beta+1). \label{eq6}
\end{eqnarray}

The cosmological evolution equations based on the theory are
\begin{eqnarray}
\rho' &=& -3 \frac{a'}{a} (\rho+p) , \label{eq7} \\
\Bigl( \frac{a'}{a} \Bigr)^2 &=& \frac{8\pi G_0 \rho a^2}{3\phi} - 
\frac{a'}{a} \frac{\phi'}{\phi}+ \frac{\omega}{6} 
\Bigl( \frac{\phi'}{\phi} \Bigr)^2, \label{eq8} 
\hspace{4.0em} \\
\phi'' + 2 \frac{a'}{a} \phi' &=& \frac{1}{2\omega+3}
\Bigl\{ 8\pi G_0 a^2(\rho - 3p) - {\phi'}^2 \frac{d \omega}{d \phi} \Bigl\}.
 \label{eq9}
\end{eqnarray}
A flat universe is assumed here. 
The prime denotes a derivative with respect to the conformal time. 
$\rho$ and $p$ are the total energy density and pressure, respectively.
The effective gravitational constant measured by Cavendish-type experiments 
is given by \cite{CMW,DE}
\begin{eqnarray}
G(\phi)={G_0\over \phi}{2\omega(\phi)+4\over 2\omega(\phi)+3}.
\end{eqnarray}
The requirement that today's gravitational constant be in agreement with
the Newton's constant determines the present value of $\phi$ as
\begin{eqnarray}
\phi_0 = \frac{4+2\omega_0}{3+2\omega_0} = 1 + {\alpha_0}^2, \label{eq10}
\end{eqnarray}
where $\omega_0$ denotes the present value of $\omega(\phi)$. 
The system of perturbation equations is found in ref.\cite{NCS}.

\begin{figure}
\includegraphics[width=\hsize]{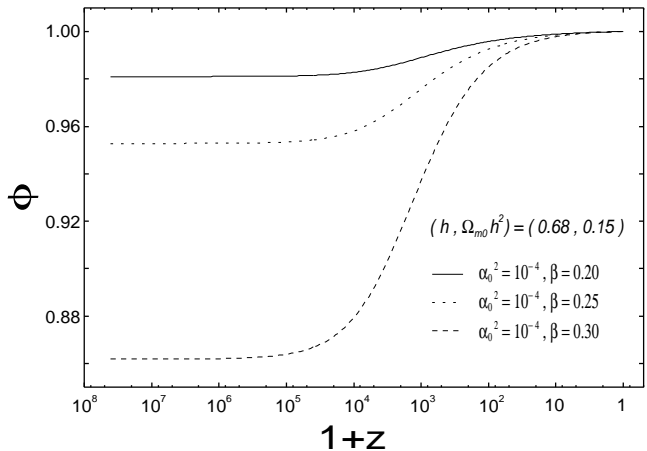}
\caption{
The typical time evolution of $\phi$ in the scalar-tensor models with
$\Lambda$CDM parameters. The smaller $\phi$ in the early epochs results in the
smaller horizon length than that in the Einstein model.
}
\label{fig:phi_evo}
\end{figure}

Fig.\ref{fig:phi_evo} shows the examples of typical $\phi$ evolution. 
$\phi$ is frozen during the radiation-dominated epoch and 
begins to grow at the matter-radiation equality time 
to realize the Newtonian gravitational constant at present. 
As $\phi$ moves upward, $2\omega+3$ increases toward infinity, and therefore 
the present small deviation from the Einstein gravity is naturally realized. 
The increase of $2\omega+3$ decelerates the motion of $\phi$ and 
finally $\phi$ converges to some value which corresponds to the minimum 
of the effective potential. As increasing $\alpha_0$ or $\beta$, we obtain 
smaller initial $\phi$. This results in the smaller horizon length in the 
early epochs, which leaves distinct traces on the CMB spectrum. 

\begin{figure}
\includegraphics[width=\hsize]{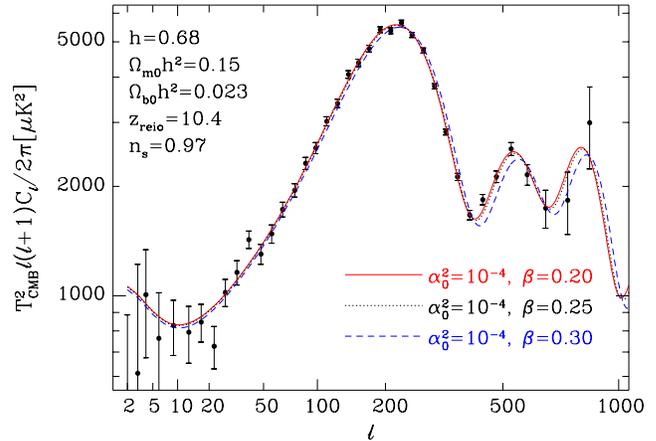}
\caption{
The typical CMB temperature anisotropy spectra in the scalar-tensor 
models with $\Lambda$CDM parameters. Also the WMAP data is 
displayed for reference. 
The normalizations are marginalized to the WMAP data. 
The acoustic peak locations are shifted to 
smaller angular scales due to the smaller sound horizon length 
(cf. Fig.\ref{fig:phi_evo}).
}
\label{fig:cmbt}
\end{figure}

The typical CMB anisotropy spectra are shown in Fig.\ref{fig:cmbt}. 
The locations of acoustic peaks (sound horizon scale) and that of diffusion 
damping tail are dependent on the horizon length at the recombination epoch. 
Since the matching condition of $\phi_0$ restricts the deviation of the
present horizon length from that in the Einstein gravity, 
these angular scales directly represent the
horizon length at recombination. Therefore, the shift of these angular 
scales to smaller scales is due to the large gravitational constant at 
recombination. The locations of acoustic peaks are shifted to 
smaller angular scales in proportion to the horizon length. Although the 
diffusion tail is also moved to smaller scale, it has weaker dependence 
on the horizon length, and hence the width between the first peak and 
the diffusion tail becomes thinner, which results in the suppression of 
small scale peaks. 
The motion of $\phi$ field and its fluctuation also distorts the 
spectrum. Roughly speaking, they make the first acoustic peak more 
prominent than higher peaks \cite{NCS}. However this effect is not 
so significant compared with the effect of changing $\Omega_{m0}$ or 
$\Omega_{b0}$.

\section{Method}

Let us describe the overview of our analysis. 
To constrain the scalar-tensor coupling parameters, 
we compare the models to the WMAP temperature anisotropy spectrum 
\cite{hins} adopting the routine provided by the authors of 
ref.\cite{verd} to compute the values of $\chi^2$. We employ a 
cosmological constant ($\Lambda$) which is the simplest dark energy 
model that can account for the late time cosmic acceleration. 

In the model concerned in this paper, the CMB spectrum depends on the three
classes of model parameters. The first class consists of the conventional 
background cosmological parameters which are today's Hubble parameter ($h$),
today's non-relativistic matter content ($\Omega_{m0}$), today's baryon
content ($\Omega_{b0}$), the redshift of cosmic reionization ($z_{reio}$),
today's CMB temperature ($T_{CMB}$), the helium mass fraction ($Y_{He}$), and
the neutrino effective number ($N_{\nu}$). We restrict ourselves to flat space
models and hence today's amount of dark energy, which is introduced as a 
component of total energy and whose equation of state is 
$p_{\Lambda} / \rho_{\Lambda} = -1$, 
is a function of $\Omega_{m0}$ so that the modified 
Friedmann equation (Eq.(\ref{eq8})) is satisfied. 
The second class consists of perturbation parameters 
characterizing the origin of fluctuations. They are the combination of 
adiabatic and isocurvature initial perturbations, the scalar spectral index 
($n_s$) and the overall normalization ($A$). 
The last class consists of the scalar-tensor 
coupling parameters which are the present deviation from the Einstein gravity 
(${\alpha_0}^2$) and the curvature ($\beta$) of 
the harmonic effective potential in the Einstein conformal frame.

For the cosmological parameters, we use the following priors that are 
principally based on the WMAP 68\% mean confidence range. 
Hubble parameter ($h$) takes the value between $0.67$ and $0.77$. 
As described in the next section, larger $\Omega_{m0}$ (smaller 
cosmological constant) shifts the acoustic peak
locations to larger angular scales also in the scalar-tensor models. 
Therefore we allow the non-relativistic matter of somewhat large
content, $\Omega_{m0} \in (0.10, 0.78)$, to confirm how small the value of 
$\chi^2$ for a set of $({\alpha_0}^2, \beta)$ can be at its minimum.
Similarly, the range of today's baryon content is set as $\Omega_{b0} 
\in (0.023,
0.055)$. We employ instantaneous reionization at the redshift ($z_{reio}$)
between 12 and 22. 
We set the CMB temperature as $T_{CMB}=2.726K$ from {\it COBE} \cite{mather}
and the neutrino effective number $(N_{\nu})$ is fixed at $3.04$. 
Although the helium mass fraction depends on the result of BBN, which is
dependent on other parameters, it does not significantly affect the CMB
spectrum. Therefore we set it as $Y_{He}=0.24$. 

According to the studies devoted to the generation of fluctuations during 
particular inflation models with scalar-tensor gravity \cite{sty,gbw,CSY}, 
it is found that, 
although isocurvature perturbations could be produced during scalar-tensor 
inflation, they
are in general negligible compared with adiabatic perturbations and then
the spectrum of the initial perturbations is not precisely scale invariant. 
Hence we employ adiabatic initial condition and allow the scalar spectral
index to take the slightly deviated value from scale invariant one as
follows: $n_s \in (0.95, 1.03)$. As the overall normalization for each model,
we survey the region between $0.8$ and $2.2$ in unit of {\it COBE} 
normalization factor.

Since we are interested in setting a constraint from CMB 
alone, we survey the ranges of $\alpha_0$ and $\beta$ including 
the regions which have been ruled out by the solar constraint
shown in Eq.(\ref{eq6}) 
as 
\begin{eqnarray}
{\alpha_0}^2 & \in & (4 \times 10^{-8}, 4 \times 10^{-2}), \\ 
\beta & \in & (0, 0.45).
\end{eqnarray}
These regions roughly correspond to 
$10 < \omega_0 <10^7$ from Eq.(\ref{eq2}).

\section{Comparison with observations}

In this section, we show the result of $\chi^2$ test comparing the harmonic
attractor model with the WMAP data. 

\begin{figure}
\includegraphics[width=\hsize]{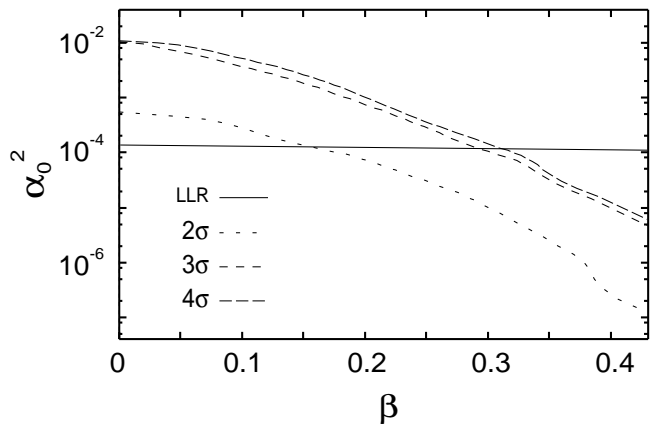}
\caption{
The $\chi^2$ contour map on ${\alpha_0}^2-\beta$ plane for the scalar-tensor
$\Lambda$CDM models, where the other parameters are marginalized. 
Also the solar bound from the LLR experiment is shown for reference.
Note that the horizontal axis ${\alpha_0}^2$ is almost proportional to 
$\gamma_{Edd} - 1$ from Eq.(\ref{eq3}). 
There exists a sharp edge approximately on the curve, $\alpha_0^2 =
10^{-2-7\beta}$. Beyond the curve, $\chi^2$ of the models rapidly increases. 
Although, in the figure, the boundary curves only up to $4 \sigma$ level are
drawn, $\chi^2$ continues to inflate for larger coupling
parameters. 
} 
\label{fig:ab}
\end{figure}

In Fig.\ref{fig:ab}, we show the $\chi^2$ contour map on 
${\alpha_0}^2-\beta$ plane, marginalizing over the other parameters. 
We find that the scalar-tensor coupling parameters are constrained as
\begin{eqnarray}
&&{\alpha_0}^2 < 5\times 10^{-4-7\beta}~~ (2\sigma),\label{eq11}\\
&&{\alpha_0}^2 < 10^{-2-7\beta}~~ (4\sigma).  \label{eq12}
\end{eqnarray}
The contour map in Fig.\ref{fig:ab} has a sharp edge 
approximately on the curve, $\alpha_0^2 = 10^{-2-7\beta}$. Beyond this curve, 
$\chi^2$ of the models rapidly increases and then those models are
statistically improbable to explain the observed spectrum. 

\begin{figure}
\includegraphics[width=\hsize]{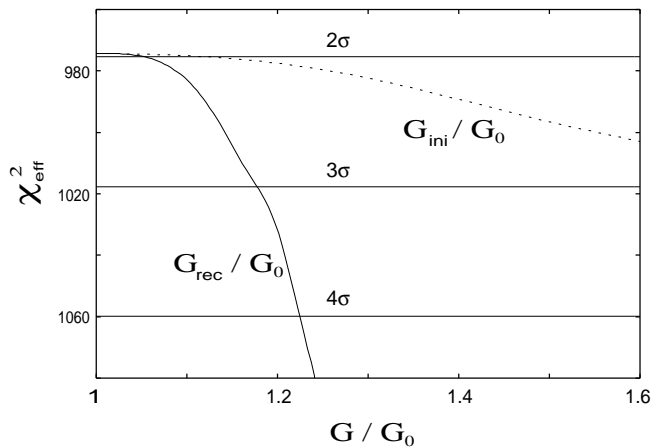}
\caption{
The $\chi^2$ for the scalar-tensor $\Lambda$CDM models as a
function of $G_{rec}$ and $G_{ini}$, where the other parameters 
are marginalized. The degree of freedom $\nu(=\sigma^2/2)$ is 891. 
}
\label{fig:grav}
\end{figure}

Next, we consider the variation of the gravitational constant 
at the recombination epoch. 
In Fig.\ref{fig:grav}, we show the $\chi^2$ for the gravitational 
constant at the recombination epoch $(G_{rec}=G(\phi_{rec}))$, 
marginalizing over other parameters. Here, $\phi_{rec}$ is the value of $\phi$ 
at the recombination epoch. 
We find that the deviation of the gravitational 
constant from today's value is constrained as
\begin{eqnarray}
&&|G_{rec}-G_0|/G_0<0.05~~(2\sigma),\\
&&|G_{rec}-G_0|/G_0<0.23~~(4\sigma).
\end{eqnarray}
The deviation of the gravitational constant 
from today's value up to 5 \% does not significantly 
change $\chi^2$ and the degree of fit of such models is comparable to that 
of WMAP team's best fit model, while even larger deviation increases $\chi^2$.
In the models on the curve, the locations of acoustic peaks are 
fitted to those of the observed spectrum and then, in order to 
pull back the peaks shifted by the larger gravitational constant 
to observed locations, $\Omega_{m0}$ and $\Omega_{b0}$
are out of their favorable values to fit the shape of the spectrum. 
The $\chi^2$ for the gravitational constant at our initial time
($G_{ini}=G(\phi_{ini}))$ is also displayed for reference. 
Here, $\phi_{ini}$ is the value of $\phi$ 
at $z_{ini}(\sim 10^8)$. 
We find that the CMB constraint 
on the gravitational constant at $z_{ini}$, $|G_{ini}-G_0|/G_0<0.12$, is 
on the same order as the BBN bound \cite{AKR}: 
$0.7<G_{BBN}/G_0<1.4 (2\sigma)$, and it can be comparable to 
the BBN bound on the harmonic model Eq.(\ref{eq2}) \cite{DP} since our 
analysis is limited to $\beta<0.45$. 

\begin{figure}
\includegraphics[width=\hsize]{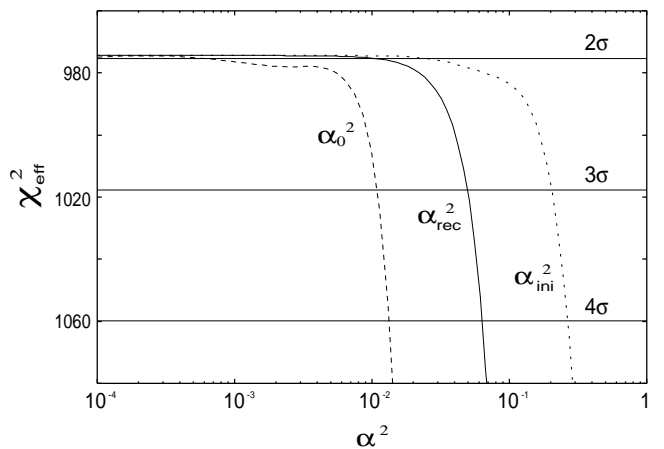}
\caption{
The $\chi^2$ for the scalar-tensor $\Lambda$CDM models as a
function of ${\alpha_0}^2$,${\alpha_{rec}}^2$ and ${\alpha_{ini}}^2$, 
where the other parameters are marginalized. 
The curve for ${\alpha_0}^2$ is idential to the $\chi^2$ for the 
Brans-Dicke models. 
The degree of freedom $\nu(=\sigma^2/2)$ is 891.
}
\label{fig:adec}
\end{figure}

In Fig.\ref{fig:adec}, we show the allowed post-Newtonian deviations 
at three different epochs. 
The curve for today's deviation parameter (${\alpha_0}^2$) is 
identical to the cross section of the contour map at $\beta=0$ and hence 
it is identical to the $\chi^2$ curve for the Brans-Dicke models. 
The sharp edge mentioned above is located around 
${\alpha_0}^2 \sim 10^{-2}  (\omega_0 \sim 50)$. 
The models whose Brans-Dicke parameter is much smaller than the lower 
bound by the solar system experiments can be compatible with CMB fluctuation, 
which was expected previously. Even if we require that the degree of fit 
should be comparable to that of WMAP team's best fit model, the boundary 
of the corresponding region is still beyond the solar bound. 
On the other hand, the deviation at the recombination epoch and $z_{ini}$ 
are relatively loosely constrained: 
${\alpha_{rec}}^2 = 1/(2\omega(\phi_{rec})+3) < 7 \times 10^{-2}, 
 {\alpha_{ini}}^2 = 1/(2\omega(\phi_{ini})+3)  < 2 \times 10^{-1} 
(4\sigma)$. This might have implications for extended inflation 
scenarios \cite{LS}-\cite{L}.

\begin{figure}
\includegraphics[width=\hsize]{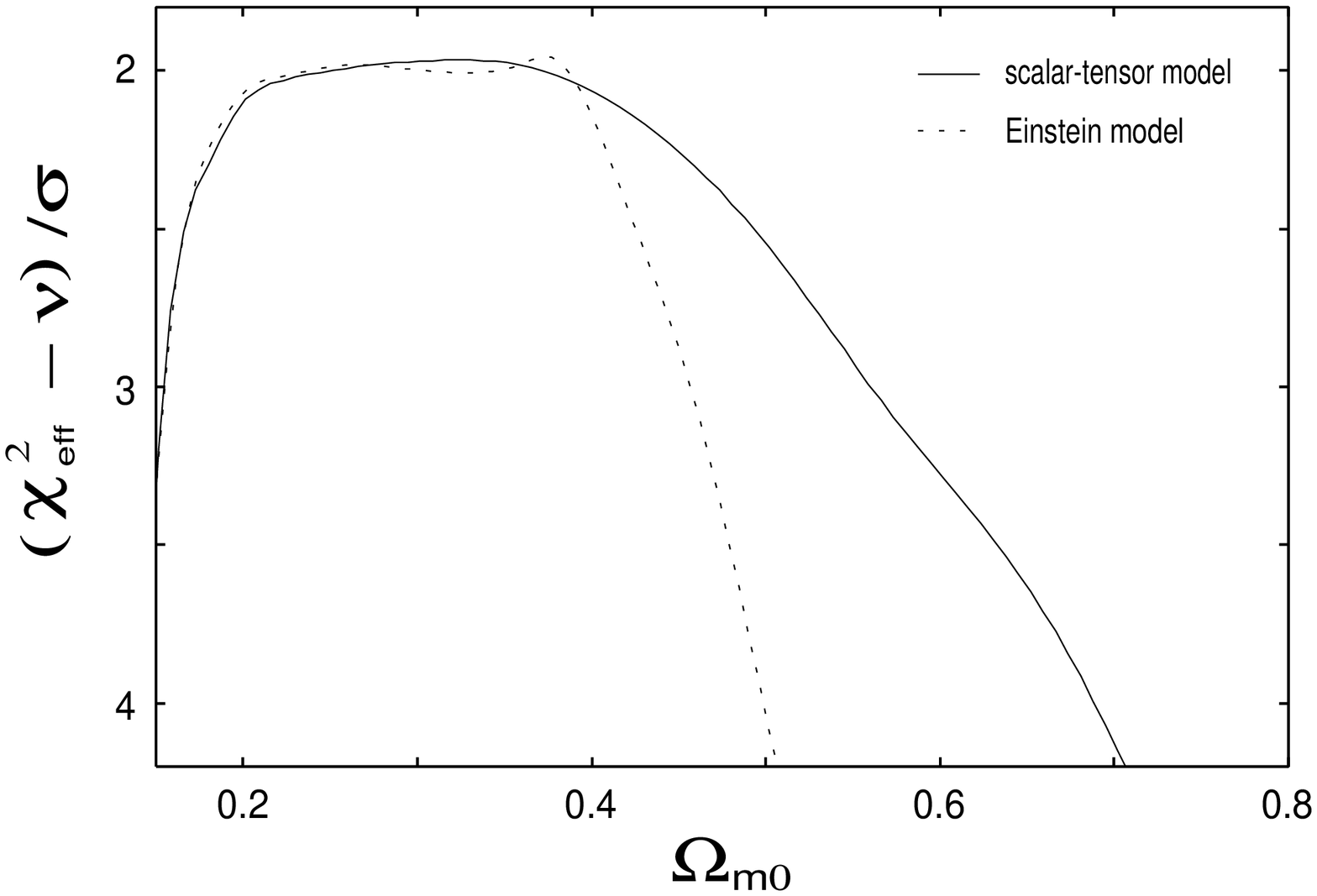}
\caption{
The $\chi^2$ for the Einstein and the scalar-tensor $\Lambda$CDM models as a
function of $\Omega_{m0}$, where the other parameters are marginalized. 
$\nu(=\sigma^2/2)$ in the scalar-tensor and the Einstein models are 891 
and 893, respectively.
}
\label{fig:omm}
\end{figure}

\begin{figure}
\includegraphics[width=\hsize]{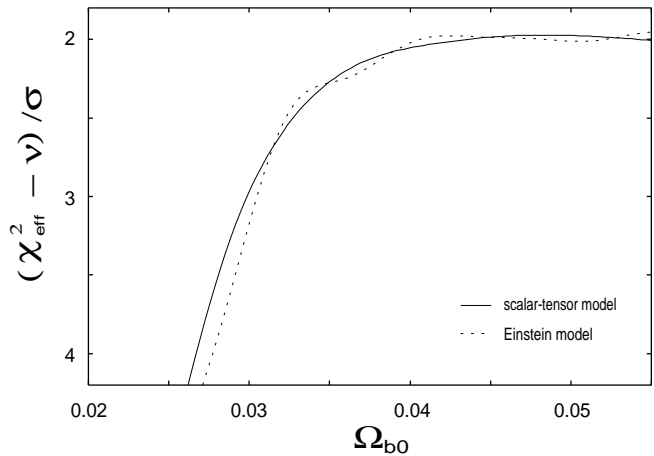}
\caption{
The $\chi^2$ for the Einstein and the scalar-tensor $\Lambda$CDM models as a
function of $\Omega_{b0}$, where the other parameters are marginalized. 
$\nu(=\sigma^2/2)$ in the scalar-tensor and the Einstein models are 
891 and 893, respectively.
}
\label{fig:omb}
\end{figure}

In Figs.\ref{fig:omm} and \ref{fig:omb}, we plot the $\chi^2$ for 
$\Omega_{m0}$ and $\Omega_{b0}$, marginalizing over other 
parameters. Compared with the Einstein models, the scalar-tensor models 
can be more probable in high density models because the scalar-tensor 
models have the two more tunable parameters to fit acoustic peak 
locations. Although the peak locations are dependent on baryon abundance, 
large amount of non-relativistic matter results in very low peak heights 
which cannot be compensated by baryon drag especially on the 2nd peak. 
On the other hand, in low $\Omega_{b0}$ models, there is no definite 
difference between the scalar-tensor and the Einstein models. 
The observed spectrum have the prominent peaks and hence low density and small
baryon abundance models are not so ill-fitted compared with the scalar-tensor
models within the surveyed parameter range.

Finally, we comment on the case if we allow non-flat universe models. 
Even in the case of non-flat models, the constraint for the scalar-tensor
coupling parameters (${\alpha_0}^2, \beta$) of Eq.(\ref{eq12}) 
would not significantly change 
because the value of $\phi$ in the early epochs depends on 
these parameters exponentially. On the other hand, the constraint for the 
gravitational constant at the recombination epoch would become much weaker 
because acoustic peak locations can be easily modulated in curved models 
without disturbing CDM or baryon abundance significantly. 
This degeneracy would be improved by observations at even smaller scales 
 uncovering the diffusion cut-off \cite{PLANCK} because 
the diffusion length depends on horizon length in a different way from 
peak locations. The ratio of the angular scale of the sound horizon 
to that of the diffusion scale is independent of projection effect, and 
hence it can provide the information on the horizon length at 
the recombination epoch. 

\section{Conclusion}

In this paper, we have quantitatively compared the CMB temperature
spectrum in the scalar-tensor flat $\Lambda$CDM model to the WMAP data by 
$\chi^2$ test of goodness of fit. 

The present deviation from the Einstein gravity ($\alpha_0^2$) must 
be smaller than $5\times 10^{-4-7\beta}$ 
($2\sigma$), and $10^{-2-7\beta}$ ($4\sigma$) for $0< \beta<0.45$. 
This constraint is much stronger 
than the bound from the solar system experiments for 
large $\beta$ models, i.e., $\beta> 0.2$ and $0.3$ in 
$2\sigma$ and $4\sigma$ limits, respectively. 
Within the framework of the harmonic attractor model, 
the difference between the 
gravitational constant at the recombination epoch and at the present is 
constrained as $|G(z=z_{rec})-G_0|/G_0 < 0.05(2\sigma)$ and 
$0.23(4\sigma)$. This is the first-time bound on the variation of the 
gravitational constant from CMB anisotropy spectrum. 
While the present deviation from the Einstein gravity is severely 
constrained ($\alpha_0^2< 10^{-4}$), larger deviation during radiation 
dominated epochs is compatible with CMB. 
Indeed ${\alpha_{ini}}^2$ up to $2 \times 10^{-2}$ can be 
within $2 \sigma$ level. Although our analysis is limited to flat models, 
the further small scale CMB anisotropy data which 
will be provided in future \cite{PLANCK} would break the degeneracy 
and would significantly improve the bound.

 \begin{acknowledgments}
T.C. and N.S. were supported in part by a Grant-in-Aid for Scientific 
Research (No.15740152 and No.14340290) 
from the Japan Society for the Promotion of
Science. T.C. was also supported by a 
Grant-in-Aid for Scientific Research on Priority Areas 
(No.14047212) from the Ministry of Education, Science, Sports and 
Culture, Japan. 
 \end{acknowledgments}

\end{document}